\documentclass{Interspeech2024}
\usepackage{cite}
\usepackage{caption}
\usepackage{subcaption}
\usepackage{url}
\usepackage{tablefootnote}
\usepackage{graphicx}
\usepackage{amsmath}
\usepackage{multirow}
\usepackage{amssymb}
\usepackage{pifont}
\usepackage{cite}
\usepackage{booktabs}
\usepackage{xcolor,tabularx}
\usepackage{hyperref}
\usepackage{lipsum}
\usepackage{tabularx,lipsum,amssymb,amsmath,pifont,graphicx,adjustbox}
\usepackage{listings}
\hypersetup{
    colorlinks=true,
    linkcolor=purple,
    filecolor=magenta,      
    urlcolor=purple,
}

\newcolumntype{Y}{>{\centering\arraybackslash}X}

\newcommand{\newpara}[1]{\vspace{2pt}\noindent\textbf{#1}}

\interspeechcameraready

\title{Can you Remove the Downstream Model for Speaker Recognition\\with Self-Supervised Speech Features?}
\name[affiliation={1}]{Zakaria}{Aldeneh}
\name[affiliation={1}]{Takuya}{Higuchi}
\name[affiliation={2}]{Jee-weon}{Jung}
\name[affiliation={1}]{Skyler}{Seto}
\name[affiliation={1}]{Tatiana}{Likhomanenko}
\name[affiliation={1}]{\\Stephen}{Shum}
\name[affiliation={1}]{Ahmed Hussen}{Abdelaziz}
\name[affiliation={2}]{Shinji}{Watanabe}
\name[affiliation={1}]{Barry-John}{Theobald}
\address{$^1$Apple, USA\qquad$^2$Carnegie Mellon University, USA}
\email{\{zaldeneh, takuya\_higuchi\}@apple.com}

\begin{document}

\maketitle

\begin{abstract}
Self-supervised features are typically used in place of filter-bank features in speaker verification models. However, these models were originally designed to ingest filter-banks as inputs, and thus, training them on self-supervised features assumes that both feature types require the same amount of learning for the task. In this work, we observe that pre-trained self-supervised speech features inherently include information required for a downstream speaker verification task, and therefore, we can simplify the downstream model without sacrificing performance. To this end, we revisit the design of the downstream model for speaker verification using self-supervised features. We show that we can simplify the model to use $97.51\%$ fewer parameters while achieving a $29.93\%$ average improvement in performance on SUPERB. Consequently, we show that the simplified downstream model is more data efficient compared to the baseline---it achieves better performance with only $60\%$ of the training data.
\end{abstract}
Self-supervised learning, representation learning, speaker recognition, speaker verification

\vspace{-5pt}
\section{Introduction}
Self-Supervised Learning (SSL) for speech (e.g., wav$2$vec~$2.0$~\cite{baevski2020wav2vec}, HuBERT~\cite{hsu2021hubert}, w2v-BERT~\cite{chung2021w2v}, BEST-RQ~\cite{chiu2022self}) enables learning powerful representations using a large amount of unlabeled data. Once trained on  unlabeled data, SSL models can be fine-tuned on labeled data to achieve remarkable performance on target downstream tasks (e.g., automatic speech recognition, speaker recognition, language identification, emotion recognition)~\cite{fan2020exploring,wang2021fine,chen2022does,chen2022large,stafylakis2023extracting,peng2023parameter}. Fine-tuning a pre-trained SSL model for each downstream task, however, can be costly due to computation and memory constraints. A more appealing setup is using an SSL model as a \textit{general-purpose feature extractor}, where the pre-trained model is frozen, and features extracted from this frozen model are used with smaller, task-dependent downstream models~\cite{yang2021superb}. In this work, we study the role the downstream model plays when using general-purpose SSL features for speaker verification.

Downstream speaker verification models (such as x-vector~\cite{snyder2018x} and ECAPA-TDNN~\cite{desplanques2020ecapa}) used in prior works were originally designed to ingest filter-bank features as inputs, whereas state-of-the-art SSL models operate on raw waveforms and the features that the SSL models extract are learned in an end-to-end manner using a Transformer architecture~\cite{vaswani2017attention,devlin2018bert,liu2019roberta}.  An important difference here is that unlike features derived from filter-banks, features from SSL models capture long-form contextual information in their representation, and it has been shown that this information is useful for predictive speech processing tasks~\cite{masumura2019large,shon2023context,zhao2020improving}.  In addition, SSL representations capture speaker information as it was shown that explicitly disentangling speaker information during SSL pre-training results in improved performance on content related tasks~\cite{qian2022contentvec}.  These findings suggest that  SSL pre-training may have already done some of the learning required for downstream speaker-related tasks, whereas models trained on top of filter-bank features must still extract all of this information from the network inputs.

Given the contrast between filter-banks and SSL features, we explore the role the downstream model plays in speaker verification with SSL features. We first seek to understand the capability of several SSL models when conducting speaker verification without a downstream model (i.e., zero-shot capability) and use the findings from our analyses to revisit the design of the downstream speaker verification model. Specifically,  we show that we can reduce the capacity of the downstream speaker verification model by $97.51\%$ and still obtain a $29.93\%$ average improvement in performance on the SUPERB~\cite{yang2021superb} benchmark. Additionally, we show that the simplified downstream model is especially effective in limited training data scenarios, outperforming its baseline counterpart with only $60\%$ of the training data. 

\begin{table*}[th]\footnotesize
  \caption{Zero-shot (i.e., no downstream model) speaker verification performance of SSL models.  ``Params.'' denotes the number of parameters in the model; ``Data'' denotes the data used for training the model; $\Delta$ denotes the relative ($\%$) improvement that SSL features provide over using filter-bank (FBank) features. The values for ``Params.'' and ``Data'' columns are taken from~\cite{yang2021superb}.}
  \label{tab:zero_shot}
\centering
\resizebox{0.9\textwidth}{!}{
\begin{tabular}{lllrrrrr}
\toprule
\multirow{2}{*}{\textbf{Model}}  &  \multirow{2}{*}{\textbf{Params.}} & \multirow{2}{*}{\textbf{Data}}  & \multicolumn{2}{c}{\textbf{LibriSpeech (in-domain)}} & \multicolumn{2}{c}{\textbf{VoxCeleb1 (out-of-domain)}} \\
 & & &  EER ($\%$)  $\downarrow$ & $\Delta$ ($\%$) $\uparrow$ &  EER ($\%$)  $\downarrow$ & $\Delta$ ($\%$)  $\uparrow$\\

\midrule
FBank &  $-$ &          $-$ &                    $7.2$ &        $0.0$ &                $40.4$ &    $0.0$ \\

\midrule
\midrule

HuBERT (base) &   $94.68$M &   LS $960$ hr &          $4.4$ &        $38.9$ &               $32.0$ &    $20.8$ \\
HuBERT (large)  &   $316.61$M &  LL $60$k hr &          $3.1$ &        $56.9$ &               $31.7$ &    $21.5$ \\

\midrule

wav2vec 2.0 (base) &   $95.04$M &   LS $960$ hr &          $5.5$ &        $23.6$ &               $33.2$ &    $17.8$ \\
wav2vec 2.0 (large)  &   $317.38$M &  LS $960$ hr &          $2.6$ &        $63.9$ &               $30.7$ &    $24.0$ \\
wav2vec 2.0 (large)  &   $317.38$M &  LL $60$k hr &          $2.8$ &        $61.1$ &               $27.2$ &    $32.7$ \\
wav2vec 2.0 (large)  &   $317.38$M &  VoxPopuli $100$k hr &  $3.1$ &        $56.9$ &               $32.7$ &    $19.1$ \\

\midrule

WavLM (base)  &   $94.70$M &   LS $960$ hr &          $4.7$ &        $34.7$ &               $31.3$ &    $22.5$ \\
WavLM (base+)  &   $94.70$M &   Mix $94$k hr$^{*}$ &         $4.0$ &        $44.4$ &               $31.3$ &    $22.5$ \\
WavLM (large)  &   $316.62$M &  Mix $94$k hr$^{*}$ &         $2.5$ &        $65.3$ &               $23.0$ &    $43.1$ \\

\midrule

wav2vec &              $32.54$M &   LS $960$ hr &          $5.3$ &        $26.4$ &               $30.7$ &    $24.0$ \\
vq-wav2vec &           $34.15$M &   LS $960$ hr &          $11.4$ &       $-58.3$ &              $37.8$ &    $6.4$ \\
Modified CPC &      $1.84$M &    LL $60$k hr &          $3.5$ &        $51.4$ &               $27.9$ &    $30.9$ \\



\bottomrule
\end{tabular}}
\\
$^{*}$ The dataset contains GigaSpeech~\cite{chen2021gigaspeech}, which includes samples collected from YouTube. 
\vspace{-12pt}
\end{table*}

\vspace{-5pt}

\section{Related Work}
In this section, we discuss relevant prior works that looked at the intersection of SSL and speaker recognition. Specifically, we focus on SSL approaches that learn \textit{generic} representations rather than approaches designed to extract specialized representations for speaker verification (e.g.,~\cite{cho2022non
,cho2020learning,zhang2021contrastive}). We refer the reader to~\cite{mohamed2022self} for a thorough review on SSL representations.


Fan et al.~\cite{fan2020exploring} studied the effectiveness of a wav$2$vec $2.0$ model on speaker verification and language identification. The authors visualized the features extracted from the model to show that the features capture speaker and language information. The authors then attached a fully-connected layer to the top of the model and ran experiments (both with and without fine-tuning the full model) to quantitatively demonstrate effectiveness of the pre-trained model on the downstream tasks. In contrast to~\cite{fan2020exploring}, our work presents a comparative study that quantifies (not visualizes) the speaker information captured by several state-of-the-art SSL models (not just wav$2$vec $2.0$). In addition, our work presents a study into the role the downstream model plays when performing speaker verification using SSL features. 

Chen et al.~\cite{chen2022does} ran experiments to understand the components of pre-training that affect the performance of SSL models when fine-tuned on the speaker verification task. The authors used a weighted average of the hidden states, which is then passed to a downstream model for learning the speaker embeddings. Their results suggested that SSL models provide better features for speaker verification compared to those extracted from a model trained on the same dataset to perform automatic speech recognition. Chen et al.~\cite{chen2022large} examined the impact of different pre-training methods, SSL model sizes, and training datasets on the downstream performance. The results reaffirmed the benefit of SSL features over filter-bank features; and the importance of data augmentation for achieving state-of-the-art performance when training the downstream model. In contrast to~\cite{chen2022does} and~\cite{chen2022large}, our experiments do not use a fixed downstream architecture; instead, we focus on re-designing the downstream model given the differences between filter-banks and SSL features.

Peng et al.~\cite{peng2023parameter} studied parameter-efficient fine-tuning to adapt pre-trained SSL models for speaker verification. They showed that using adaptors is better than fine-tuning the full SSL model in low-resource settings. Stafylakis et al.~\cite{stafylakis2023extracting} proposed correlation pooling, an approach for aggregating frame-level SSL features across time to induce fixed-size utterance-level features. The authors showed that replacing statistics pooling  with correlation pooling improved the performance of speaker verification when using SSL features. In contrast to~\cite{peng2023parameter} and ~\cite{stafylakis2023extracting}, our work does not study adaptors (i.e., the addition of modules between the layers of the pre-trained SSL model)---we use SSL models as generic feature extractors and focus on the design of the full downstream model, not just the pooling mechanism.

\begin{table}[t!]
  \caption{An unconstrained speaker verification setup yields a lower equal error rate (EER, $\%$) on VoxCeleb1 compared to SUPERB. We fine-tune both the WavLM and ECAPA-TDNN models for the unconstrained setup; we include the VoxCeleb2 during fine-tuning and apply training-time augmentations.}
  \label{tab:sota}
  \centering
  \begin{tabularx}{\linewidth}{lYY}
    \toprule
    \textbf{Model} & \textbf{SUPERB} & \textbf{Unconstrained}\\
    \midrule
    WavLM+ECAPA-TDNN & $2.03$ & $0.39$ \\
    \bottomrule
  \end{tabularx}

  \vspace{-15pt}
\end{table}

\vspace{-5pt}
\section{Experiments}

This work explores the design of the downstream speaker verification architecture given the contrasting nature of filter-banks and SSL features. We begin with an investigation into the speaker information that is captured by state-of-the-art SSL methods (Section~\ref{subsec1}). We then run an ablation on the downstream architecture to understand the role different components play in speaker verification using SSL features (Section~\ref{subsec2}).

\begin{table*}[th]\footnotesize
  \caption{We can simplify the downstream model for speaker verification when using SSL features. The ``base'' SSL models were trained on LibriSpeech $960$ hr, and the ``large'' SSL models were trained on Libri-Light $60$k hr. The downstream models were trained on the development set of VoxCeleb$1$. The equal error rates (EERs, $\%$) are reported on the verification set of VoxCeleb$1$.}
  \label{tab:downstream_ablation}
\centering
\resizebox{\textwidth}{!}{
\begin{tabular}{llcrccccccc}
\toprule
& \multirow{3}{*}{\textbf{Downstream}}  & \multirow{3}{*}{\textbf{Params.}} & & \multicolumn{7}{c}{\textbf{EER (\%) $\downarrow$}} \\
& & &\multirow{2}{*}{FBank} & HuBERT  & wav2vec 2.0 & WavLM & & HuBERT  & wav2vec 2.0 & WavLM \\
& & & &  base  & base & base & & large  & large & large \\

\midrule

A & x-vector~\cite{yang2021superb, chen2022wavlm}& $5.7$M &     $9.56$ &   $5.11$ &       $6.02$ &           $4.69$ &    &    $5.98$ &        $5.65$        & $3.77$  \\

B & x-vector (our setup)                                & $5.7$M &     $9.95$ &   $4.41$ &       $5.12$ &           $4.97$ &    &    $5.28$ &        $6.04$ &      $4.04$ \\
C & Stafylakis et al.~\cite{stafylakis2023extracting}                                & $-$ &     $-$ &   $-$ &       $-$ &           $-$ &    &    $4.8$ &        $-$ &      $3.8$ \\

D.$0$ & ECAPA-TDNN                                          & $8$M &     $8.78$ &   $3.81$ &       $4.63$ &           $3.67$ &    &    $3.14$ &        $6.35$ &      $2.03$ \\

 & \quad $\rightarrow$ w/o frame-level encoder                       & &   &  &    &       &    &    &         &     \\

D.$1$ & \quad\quad $\rightarrow$ w/ channel \& context stats. pooling                       & $725$K &     $8.75$ &   $2.95$ &       $3.41$ &           $2.71$ &    &    $2.46$ &        $2.60$ &      $1.55$ \\


D.$2$ & \quad\quad $\rightarrow$ w/ attentive stats. pooling                     & $725$K &     $9.26$ &   $2.87$ &       $3.09$ &          $2.51$ &    &    $2.37$ &        $2.56$ &      $1.70$\\


D.$3$ & \quad\quad $\rightarrow$ w/ stats. pooling                   & $199$K &     $10.74$ &  $3.11$ &       $3.19$ &           $2.83$ &    &    $2.34$ &        $2.63$ &      $1.56$\\
\midrule

& $\Delta (\text{D.$0$}, \text{D.$3$})$ ($\%$)  $\uparrow$     & $97.51$ &    $-22.32$ & $18.37$ & $31.10$ & $22.89$ & & $25.48$ & $58.58$ & $23.15$ \\

\bottomrule
\end{tabular}}
\vspace{-15pt}
\end{table*}

\subsection{What Speaker Information is Captured by SSL?}\label{subsec1}

\newpara{Motivation.}
Results from SUPERB~\cite{yang2021superb} suggest that SSL models capture speaker information even though these models were trained with frame-level objectives that resemble the objective of automatic speech recognition. The captured speaker information can be undesirable if the goal is to learn models that focus on content related tasks~\cite{qian2022contentvec}. We identified two limitations in prior analyses that we address in our work. First, prior analyses focused only on reporting the speaker verification performances of SSL models on the VoxCeleb1 dataset~\cite{nagrani2020voxceleb}. However, VoxCeleb1 is an out-of-domain data for several SSL models that are evaluated in the literature---these SSL models were trained on audiobooks. Thus, it is unclear if the performance degradation comes from out-of-domain data, or from the task itself. Second, prior analyses used SSL features along with either x-vector or ECAPA-TDNN architectures for the verification task. However, there is evidence suggesting that the performance and the ranking is highly sensitive to the choice of the downstream model~\cite{zaiem2023speech}. To this end, we seek to understand the capability of SSL models to do speaker verification \textit{without a downstream model} (i.e., \textit{zero-shot capability}) on both in-domain and out-of-domain data.

\newpara{Approach.}
We extract frame-level features, $\textbf{H}^l = \{\textbf{h}_{1}^{l}, \textbf{h}_{2}^{l}, \dots, \textbf{h}_{T}^{l} \}$, from layer $l \in \{1, \dots, L\}$, and then induce a fixed-dimensional representation by aggregating all $T$ frame-level features by computing the mean and standard deviation across time; where $L$ is the number of layers in the SSL model and $T$ is the number of frames in the representation. We use the cosine score to measure the similarity for trial pairs.

\newpara{Setup.}
We assess the zero-shot speaker verification capability of SSL methods on LibriSpeech (in-domain) and VoxCeleb1 (out-of-domain). We use the Vox1-O evaluation protocol and the \texttt{test-clean} and \texttt{test-other} sets of  LibriSpeech. We create a verification split for LibriSpeech by: (1) sampling all utterances that are $8<x<12$ seconds; (2) creating a list of all possible pairs; and (3) down-sampling the negative class samples such that we retain a $1$:$5$ positive-to-negative ratio in the trial list. We follow above process separately for the \texttt{test-clean} and \texttt{test-other} sets of LibriSpeech and then merge the two to obtain a list with $73$ speakers and $1908$ pairs.

\newpara{Results.}
The zero-shot speaker verification capability for several SSL models is shown in Table~\ref{tab:zero_shot}. We use zero-shot performance on filter-bank features as our baseline and discuss our findings below.

\textbf{Do SSL features highlight speaker characteristics beyond what filter-bank features highlight?} 
Features from all pre-trained SSL models (except vq-wav2vec on in-domain test set) provide improvements over filter-bank features. WavLM (large) features improve the baseline performance by $65.3\%$ and $43.1\%$ on LibriSpeech and VoxCeleb1, respectively.
Even though vq-wav2vec features degrade the baseline performance by $58.3\%$ on LibriSpeech, it outperforms the baseline by $6.4\%$ on VoxCeleb1. This result shows that SSL features can provide an improvement over filter-bank features in the zero-shot setting. Furthermore, the results reaffirm that the augmentation strategy used in WavLM training effectively  incorporates more speaker information in the representation.

\textbf{Does the domain mis-match impact the relative improvements from SSL features compared to filter-banks?} SUPERB evaluates the quality of speaker verification models on the VoxCeleb1 dataset. However, several of the SSL models are trained on audiobook data. Our results show that we obtain higher relative improvements on LibriSpeech compared to VoxCeleb1. We compute Spearman's $\rho$ between the relative improvements on the two datasets and obtain $0.66$ ($p=0.019$). This result suggests that, while there is a strong correlation between the two domains, the rankings of the SSL models are not identical and they depend on the domain of the downstream data (even for the same task).

\textbf{Are bigger SSL models better at capturing speaker information compared to smaller models?} Increasing the capacity of wav$2$vec $2.0$ from $95.04$M to $317.38$M while using the same $960$ hours of training data increases the relative improvements from $23.6\%$ to $63.9\%$ and from $17.8\%$ to $24.0\%$ on LibriSpeech and VoxCeleb1, respectively. This finding is also true for WavLM, where increasing the capacity from $94.70$M to $316.62$M while using the same $94$k hours of data increases the relative improvements from $44.4\%$ to $65.3\%$ and from $22.5\%$ to $43.1\%$ on LibriSpeech and VoxCeleb1, respectively. Our results suggest that increasing the model size provides the model with capacity to capture more information, including speaker information.

\break
\textbf{Is the prior from SSL model architecture adequate for capturing speaker characteristics?} In the vision domain, Ulyanov et al.~\cite{ulyanov2018deep} showed that the structure of convolutional networks provides a strong prior for learning. We ask whether or not SSL speech model architecture alone (i.e., no learning) provides appropriate priors for capturing speaker information from raw waveforms. We find that random SSL models, on average, drop the performance by $131.6\%$ for the LibriSpeech setup compared to baseline and drop the performance by $8.4\%$ for the VoxCeleb1 setup, suggesting that the architecture alone is insufficient for capturing speaker characteristics.

\subsection{Can we Simplify the Downstream Model?}\label{subsec2}
\newpara{Motivation.}
The results from Section~\ref{subsec1} reaffirm that SSL models capture speaker information beyond what is captured with filter-bank features. This finding suggests that we may need a different model for downstream task because the original downstream models were designed for filter-banks. We re-visit the ECAPA-TDNN architecture~\cite{desplanques2020ecapa}, noting the impact of its components on the downstream speaker verification task when used with SSL features, and propose a simple yet effective speaker verification architecture suitable for SSL features.

\newpara{Downstream Model.}
The ECAPA-TDNN model introduced enhancements to the Time Delay Neural Network (TDNN) architecture and the attentive statistics pooling layer to address the local spatial modeling limitations of convolutional networks. We study the utility of the frame-level encoder and the pooling mechanism when used with SSL features.

\textbf{Frame-level Encoder.} The frame-level encoder includes Res2Net blocks, Squeeze-Excitation (SE) blocks, and Multi-layer Feature Aggregation (MFA). The SE blocks were introduced to the the ECAPA-TDNN architecture to ``rescale the frame-level features given global properties of the recording''~\cite{desplanques2020ecapa}. MFA was introduced so the model can exploit information from multiple layers before pooling.

\textbf{Channel- and Context-dependent Statistics Pooling.} The ECAPA-TDNN architecture extends the temporal attention statistics from~\cite{okabe2018attentive} to also depend on the channel dimension. This change allows  the model to attend to different time frames for different features. The attention module calculates a scalar score for each frame given the channel: $z_{t,c} = f_{c}(\textbf{h}_t)$, where $\textbf{h}_t$ are the hidden states from the previous layer at time $t$; and $f_{c}$ is the channel-dependent non-linear transformation. The scalar scores are  normalized across time per channel. The temporal context of attentive pooling is also extended by concatenating the global context features.
 
The architectural enhancements that were introduced in downstream models can be unnecessary when these models are used with SSL features. State-of-the-art SSL models use the transformer model in their architecture. The output at each frame from a transformer already captures the full context of the utterance and the multi-head attention mechanism enables each head to focus on different aspects of the utterance. To this end, we ablate the ECAPA-TDNN architecture's structure to study the enhancements' impact on the overall performance when using SSL features.

\newpara{Setup.}
We follow the SUPERB setup for speaker verification using SSL features. We pass a waveform through a frozen SSL model and take the weighted sum of the hidden states from each layer to produce the output sequence: $\textbf{o}_t=\sum_{l=1}^{L} w^l\cdot \textbf{h}_{t}^l$, where $\textbf{h}_{t}^l$ are the hidden states from layer $l$ at time $t$; and $w^l$ is the normalized scalar weight for layer $l$. The output sequence is then fed into a downstream ECAPA-TDNN model to produce the embeddings: $\textbf{e}=\text{ECAPA-TDNN}(\textbf{o}_1, \textbf{o}_2, \dots, \textbf{o}_T)$. The summation weights and the downstream model parameters are trained to classify speakers with an additive margin softmax~\cite{wang2018additive} loss, where a scale of $30$ and a margin of $0.4$ are used. We use $512$ channels for the convolutional frame layers in ECAPA-TDNN, and use the following hyper-parameters: $\text{optim}=\text{AdamW}$~\cite{loshchilov2017decoupled}; $\text{lr}=5.0e^{-5}$; $\text{batch size}=40$. We train the models for $100\,000$ steps and create a checkpoint every $5000$ steps. We report the best performance from all checkpoints in accordance with SUPERB.

\newpara{Note.} Our work aims to study how well SSL models capture speaker information and how to extract this information effectively from frozen SSL models. Our goal is \emph{not} to achieve the best possible speaker verification performance---a goal achievable by fine-tuning the SSL model on the downstream task. We run an experiment to highlight the difference in performance on VoxCeleb1 when fixing the SSL model (WavLM) according to SUPERB and when jointly fine-tuning both WavLM and downstream models. Table~\ref{tab:sota} shows that we can achieve an EER of $0.39\%$ when the SSL and downstream models are fine-tuned simultaneously, highlighting the performance differences due to the SUPERB setup. Despite this difference in performance, we use the SUPERB setup because it is a widely used setup for benchmarking the quality of SSL features, and our focus is on limited training data scenarios for the downstream task.

\newpara{Results.}
The results of our downstream ablation are reported in Table~\ref{tab:downstream_ablation} and we discuss our findings below.

First, we replicate the x-vector setups reported in prior works to ensure a fair comparison. Our x-vector setups provide, on average, a $2.32\%$ relative improvement in performance compared to x-vector performance reported in~\cite{yang2021superb,chen2022wavlm}. This result establishes that our setup is competitive and reflects state-of-the-art performance on SUPERB. Replacing the x-vector with the ECAPA-TDNN model improves the performance by $11.76\%$ given the  filter-bank features and $22.41\%$ on average for the SSL setups, reaffirming the utility of the structural enhancements employed in ECAPA-TDNN.

We remove the frame-level encoder from the architecture and evaluate three different pooling mechanisms: channel- and context-dependent statistics pooling proposed with ECAPA-TDNN model, attentive statistics pooling from~\cite{okabe2018attentive}, and statistics pooling from~\cite{snyder2018x}. Removing the frame-level encoder reduces the model's capability to use features from multiple layers and some of its capability to exploit contextual information. ECAPA-TDNN's channel-attentive pooling mechanism without the frame-level encoder improves filter-bank performance by $0.34\%$ but improves the average SSL performance by $29.91\%$. This result suggests that SSL models do not require the same frame-level processing that filter-banks require for extracting speaker information.


Replacing channel- and context-dependent statistics pooling with attentive statistics pooling from~\cite{okabe2018attentive} drops the performance by $5.83\%$ for the filter-bank model but improves the average performance by $2.5\%$ for the SSL models. This result suggests that channel-attention is important when using filter-bank features but not when using SSL features. Finally, replacing the attentive statistics pooling with non-weighted statistics pooling reduces filter-bank performance by $15.98\%$ and reduces the average SSL performance by $2.93\%$, suggesting that the attention mechanism is less important for SSL models compared to filter-banks for speaker verification.

\begin{figure}[t]
\includegraphics[width=0.9\columnwidth]{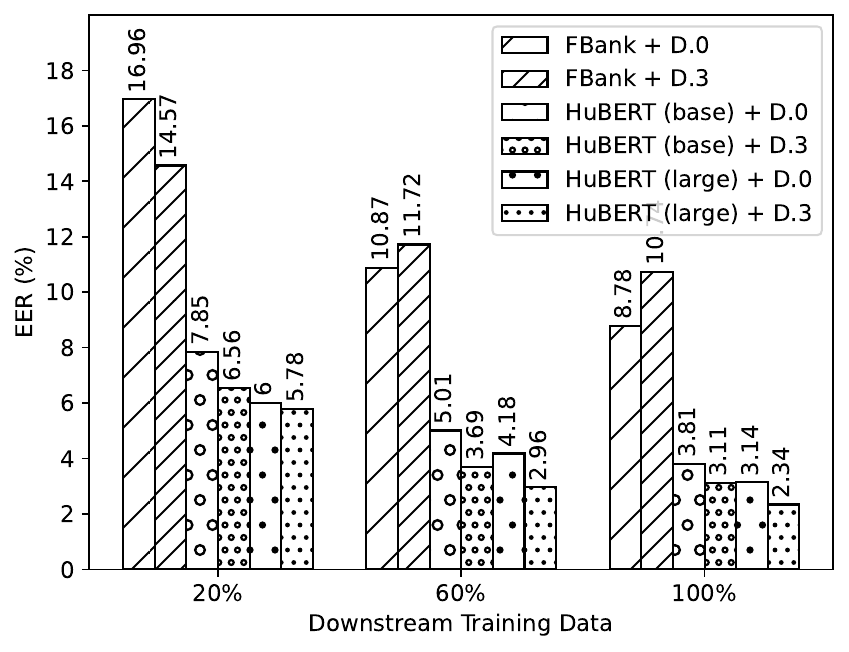}
\centering
\caption{The simplified downstream model (D.3 from Table~\ref{tab:downstream_ablation}) performs better with less data compared to full downstream model (D.0 from Table~\ref{tab:downstream_ablation}). The equal error rate (EER, $\%$) on VoxCeleb1 is reported under three data conditions: $20\%$, $60\%$, and $100\%$ of the training speakers.}
\label{data_efficiency}

\vspace{-15pt}
\end{figure}

\newpara{Data Efficiency.} We further create a random subset of the VoxCeleb$1$ containing only $20\%$ of the training speakers to study the data efficiency of the simplified downstream model similar to \cite{heo2023curriculum}. Then, we gradually increase the number of training speakers by adding randomly selected speakers until we cover $100\%$ of the data. We evaluate the downstream models using three features: filter-banks, HuBERT (base), and HuBERT (large). We focus on HuBERT in our analysis because it is widely used and it is used for pre-training WavLM.
The results in Figure~\ref{data_efficiency} show that the simplified downstream model (D.3 from Table~\ref{tab:downstream_ablation}) is more data efficient; the model achieves better performance when using only $60\%$ of the data for both HuBERT setups.

\vspace{-5pt}
\section{Conclusion}
We observed that state-of-the-art SSL features used in downstream speaker verification models were designed originally for filter-bank features. We hypothesized that downstream models can be simplified because SSL models have potentially done some of the learning required for the task. Our results suggest that, although we can't completely remove the downstream model when using SSL features, we can simplify the model to use $97.51\%$ fewer parameters and obtain a $29.93\%$ average improvement in performance compared to the original model on SUPERB. We also showed that the simplified downstream model requires less training data---the model uses $60\%$ of the original data to achieve the same or better performance compared to the full model.


\newpage





\bibliographystyle{IEEEtran}

\bibliography{mybib}

\end{document}